\documentclass{aa}  

\usepackage{graphicx}
\usepackage{xcolor}
\usepackage{soul}
\usepackage{gensymb}
\usepackage{upgreek}
\usepackage{amsmath}
\usepackage{textcomp}
\usepackage{txfonts}
\begin{document}

   \title{Single-photon gig in Betelgeuse's occultation}


   \author{F. Prada \inst{1}
          \and
          R. Gomez-Merchan\inst{2}
          \and
          E. P\'erez \inst{1}
          \and
          J. E. Betancort-Rijo \inst{3,4}
          \and
          J. A. Leñero-Bardallo \inst{2}
          \and
          \'A. Rodr\'iguez-V\'azquez \inst{2} 
          \and  \\
          G. Glez-de-Rivera \inst{5}
          \and
          S. D\'iaz-L\'opez   \inst{6,7}
          \and
          J. de Elias Cantalapiedra \inst{7}
          }

   \institute{Instituto de Astrof\'isica de Andaluc\'ia (IAA-CSIC),
              Glorieta de la Astronom\'ia s/n, E-18008 Granada, Spain\\
              \email{f.prada@csic.es}
          \and
             Instituto de Microelectrónica de Sevilla, IMSE-CNM (CSIC, Universidad de Sevilla),
             C/ Am\'erico Vespucio 28, E-41092 Sevilla, Spain
          \and
             Instituto de Astrof\'isica de Canarias, E-38205 La Laguna, Santa Cruz de Tenerife, Spain
          \and
             Departamento de Astrof\'isica, Universidad de La Laguna, E-38206 La Laguna, Tenerife, Spain
          \and 
             Grupo de Investigaci\'on HCTLab, Escuela Polit\'ecnica Superior, Universidad Aut\'onoma de Madrid, E-28049 Madrid, Spain
          \and
             Observatorio AstroCamp, Nerpio, Albacete, Spain
          \and
             Agrupación Astronómica de Madrid, C/ Albendiego 22, E-28029 Madrid, Spain
             }

   \date{...; ...}

 
  \abstract
{We present results from the occultation of Betelgeuse by asteroid (319) Leona on December 12, 2023, observed using a $64 \times 64$ pixel Single-Photon Avalanche Diode (SPAD) array mounted on a 10-inch telescope at the AstroCamp Observatory in Nerpio, Southeast of Spain, just a few kilometers from the center of the occultation shadow path. This study highlights remarkable advancements in applying SPAD technology in astronomy. The SPAD array's asynchronous readout capacity and photon-counting timestamp mode enabled a temporal resolution of 1 microsecond in our light curve observations of Betelgeuse. Our data analysis addressed challenges inherent to SPAD arrays, such as optical cross-talk and afterpulses, which typically cause the photon statistics to deviate from a Poisson distribution. By adopting a generalized negative binomial distribution for photon statistics, we accurately describe the observational data. This method yielded an optical cross-talk estimation of $1.07\%$ in our SPAD array and confirmed a negligible impact of spurious detected events due to afterpulses. The meticulous statistical examination of photon data underscores our SPAD-array's exceptional performance in conducting precise astronomical observations. The observations revealed a major decrease in Betelgeuse's intensity by $77.78\%$ at the occultation's peak, allowing us to measure Betelgeuse's angular diameter at 57.26 mas in the SDSS g-band. This measurement, employing a simplified occultation model and considering the known properties of Leona, demonstrates the potential of SPAD technology for astronomy and sets a new standard for observing ultra-rapid transient celestial events, providing a valuable public dataset for the astronomical community.}

   \keywords{Instrumentation: detectors -- occultations --
                minor planets, asteroids: individual: (319) Leona – stars: individual: Betelgeuse.
               }

   \maketitle
%

\section{Introduction}

In the dynamic realm of astronomical exploration, cutting-edge technologies such as Single-Photon Avalanche Diode (SPAD) sensors are crucial for unveiling new discoveries \cite[see][for a review]{Barbieri}. SPAD sensors excel in detecting individual photons with extraordinary timing resolution, reaching picoseconds in recent  years \cite[see][for a review]{Bronzi, Acconcia, Hadfield}. Advances have led to SPAD devices with low dark count rates (DCR) and high photon detection probabilities (PDP) \citep[e.g.][]{Vornicu,Cusini}.

However, as evidenced by prior studies \cite[e.g.][]{Zampieri2015,Zampieri2021,Matthews,Weiss,White}, the current use of single-pixel SPAD devices in astronomy and other research fields imposes a substantial limitation on their potential for spatially-resolved imaging and spectroscopy in astronomical observations. The emergence of two-dimensional SPAD detectors, however, is transforming this scenario by
enabling imaging and spectroscopic observations of transient celestial phenomena in the sub-microsecond to picosecond time domain. These advanced SPADs outperform CCD, CMOS, and IR detectors in time resolution, allowing for a deeper understanding of ultra-rapid astronomical events, capturing rapid fluctuations and fostering a deeper understanding of extreme
physical processes, and enabling applications like intensity interferometry spectroscopy at visible wavelengths to measure stellar and
black-hole accretion disk sizes \cite[see][for a review]{Barbieri,Eisenhauer}.

The growing demand for two-dimensional SPAD detectors has emerged
in recent years as a next-generation imaging technology. These devices provide millimeter-resolution depth images for future autonomous systems \cite[e.g.][]{Morimoto,Wayne,Erdogan,Ruben,Huang}. See \cite{Scholes} for a detailed review. They offer a transformative opportunity for astronomy and scientific instrumentation at large. In this context, our exploration involves testing a $64\times64$ pixel SPAD-array for astronomical observations, developed by our colleagues from the Intelligent Interface and Sensory Systems (I2CASS) group at the Instituto de Microelectr\'onica de Sevilla (IMSE CSIC-US), see \citet{Ruben}. The array is at a Technical Readiness Level (TRL) of 3. We focus on observing the extraordinary event of the occultation of the bright star $\alpha$ Orionis (Betelgeuse) by the asteroid (319) Leona on December 12, 2023. This observation provides a unique opportunity to determine Betelgeuse's light curve with remarkable time resolution. The dataset allow us to study Betelgeuse's photosphere,  precisely measuring its diameter and brightness distribution at visible wavelengths, considering Betelgeuse and Leona's expected similar angular sizes at the time of the occultation. 

The heightened interest in Betelgeuse in recent years is due to a substantial drop in brightness in the visible domain, known as the Great Dimming. This dimming is attributed to a dusty veil, indicative of vigorous activity in the stellar atmosphere of this red supergiant star \citep[see][]{Montarges,Drevon}. In preparation for Betelgeuse's occultation, the work reported by \cite{Ortiz} provides accurate dimensions and an approximate shape of the asteroid (319) Leona, determined from a previous stellar occultation.

In this letter, we present single-photon observations of Betelgeuse's occultation using our IMSE-I2CASS SPAD-array based camera attached to a 10-inch Ritchey-Chr\'etien telescope at the AstroCamp observatory in Nerpio, Spain. Section 2 provides a detailed description of the instrument, and the observations and data reduction are outlined in Section 3. The results and conclusions are presented in Sections 4 and 5, respectively.

\section{Instrument description}

Our telescope's auxiliary SPAD-camera includes a Baader-FlipMirror II Star Diagonal device, which features a precisely surfaced-mirrored flip mirror with multi-layer aluminium coating. This setup enables switching between a straight light path (Position A) for our SPAD-based camera and an angled light path (Position B) for an eyepiece graticule. The SPAD-based camera is mounted on a robotic Ritchey-Chr\'etien telescope with 10-inch aperture and f/8 focal ratio, boasting a $99\%$ dielectric high-reflectivity coating on both primary and secondary mirrors. A 2-inch SDSS$-g'$ photometric filter is integrated into the telescope's optical path to the camera. Absolute time synchronisation is achieved through a stratum-0 server with GPS Pulse Per Second (PPS). We record the GPS PPS signal only at the beginning and end of our data collection to convert the relative timestamps of photons to absolute time. These timestamps were obtained by a 100.8 MHz clock within the camera's electronics. The camera's field-of-view covers  $163.3^{\prime\prime} \times 163.3^{\prime\prime}$ on the sky, given the plate scale of $9.6\, \mu \text{m}/\text{arcsec}$ on the focal plane.

Table~\ref{tabspad} outlines the main properties of our SPAD-based camera. Each of the $64\times64$ pixels in the array includes a SPAD photodetector with associated quenching and recharge circuitry and a programmable 8-bit counter to trigger an event upon the detection of a programmable number of photons. These events are transmitted asynchronously to an external receiver using the Address Event Representation (AER) communication protocol \citep{Boahen}. The photosensitive area of each pixel is $21\, \upmu \rm m^2$, resulting in a fill factor of $3.5\%$. 
A micro-photograph of the SPAD-Array is featured in Figure 2 of \cite{Ruben}, where the system's constitutive
blocks are highlighted. The SPAD-Array chip is mounted on a PCB with peripheral circuitry dedicated to biasing the sensor event data, generating its control signals, and hosting an FPGA for processing the sensor event data. Note that the nominal time resolution of the SPAD-array is 3 ns. However, for the Betelgeuse observation, to prevent counter overflow due to the expected high rate of events and to avoid losing the time reference, we limited the time resolution to $1 \upmu \rm s$.

When a single pixel in our SPAD detector detects a user-configurable number of photons, the pixel coordinates ($x, y$) are transmitted off-chip, signaling this event. The advantages of our SPAD detector over existing technology include: 1) The user can detect the arrival of
a single photon or a packet of a programmable number of photons; 
2) Pixel information is conveyed off-chip asynchronously when the programmed number of photons per pixel has been detected. This event-driven approach eliminates the need to scan each pixel individually, thus optimizing bandwidth use and enabling very fast data acquisition with enhanced time resolution than conventional synchronous solutions. This feture is particularly advantageous for intensity interferometry observations among multiple telescopes; 
3) The arrival of individual photons or photon packets can be timestamped with an external FPGA device, adding versatility for scientific experiments; 4) At the device level, SPAD pixels feature low DCR, high
PDP at visible wavelengths, and zero read-out noise; these are critical parameters for astronomy applications.

   \begin{table}
      \caption[]{Properties of the IMSE-I2CASS SPAD-Based Camera setup for Betelgeuse observations.}
         \label{tabspad}
     $$ 
         \begin{array}{p{0.5\linewidth}l}
            \hline
            \noalign{\smallskip}
            Specification      &  Value \\
            \noalign{\smallskip}
            \hline
            \noalign{\smallskip}
             Technology     & LFoundry \, $110 nm$     \\
             Pixels         & 64 \times 64             \\
             Readout        & Asynchronous \, (event \, driven)  \\
             Pixel pitch    &   24.5\, \upmu \rm m           \\
             Fill factor    &  $3.5\%$          \\
             Spectral range &  $400 - 800 nm$            \\
             PDP            &   $75 \% @ 425 nm$         \\
             Median DCR & $13 Hz @ 12.7 \degree C$           \\
             Operation mode  &   Photon \, timestamping         \\
             Power consumption  &    $20.5 mW $@$ 10 Meps$        \\
             Max. output event rate  &   $40 Meps$         \\
             Time resolution  &   1\, \upmu \rm s         \\
            \noalign{\smallskip}
            \hline
         \end{array}
     $$ 
   \end{table}
%

\section{Observations and data reduction}

\begin{figure*}
   \centering
   \includegraphics[scale=0.67]{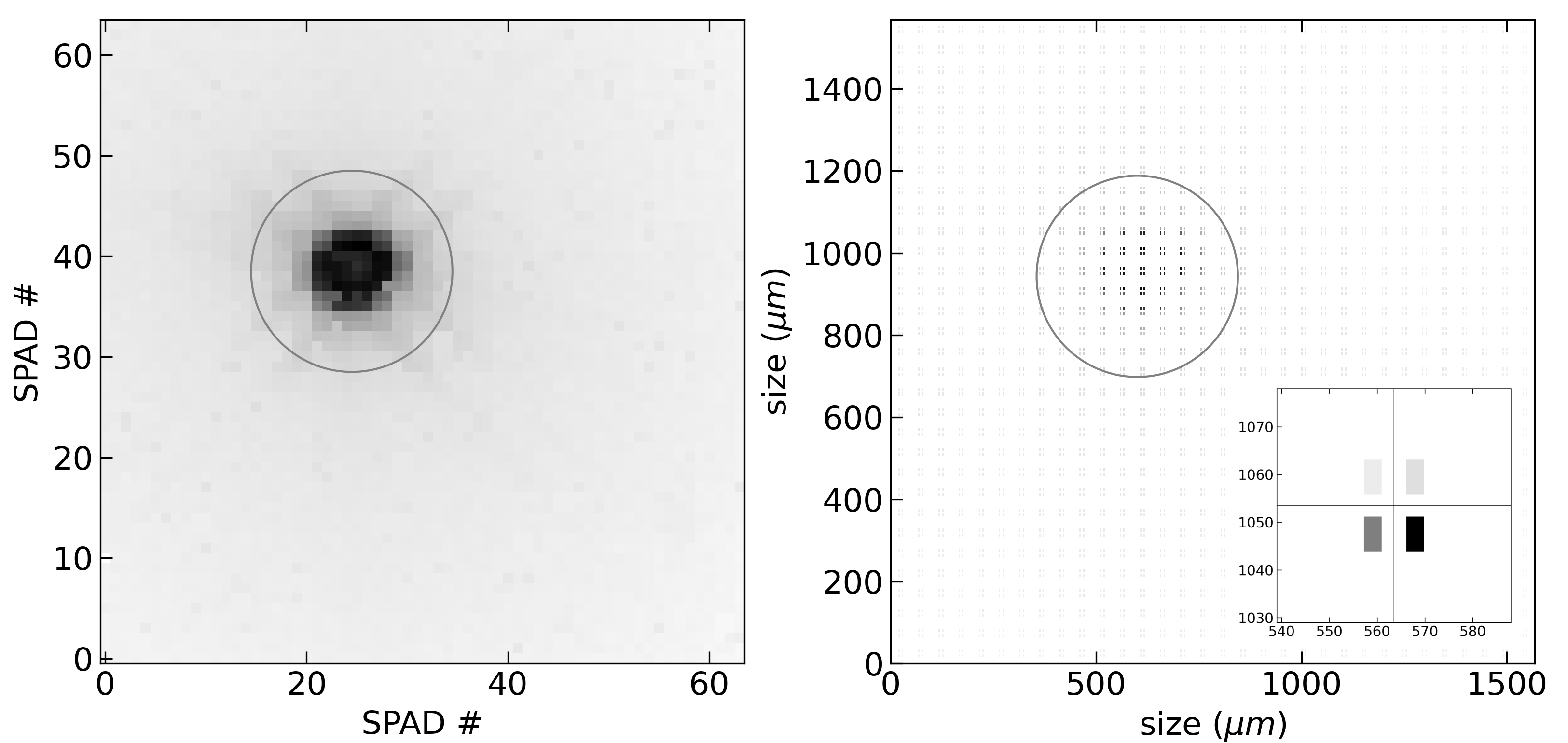}
   \caption{Defocused image of Betelgeuse during a single-chord occultation on December 12, 2023, captured with our SPAD camera from the AstroCamp observatory in Nerpio, Southeast Spain. The right panel displays a $64 \times 64$ pixel image, showcasing the SPAD pixel arrangement. The inset provides a closer view of the $2 \times 2$ SPAD arrangement. The left panel presents a symbolic image created from the $64 \times 64$ SPAD sensors, illustrating Betelgeuse's defocus image and diffraction features. Our analysis involves counting photons within a circular aperture that avoids fainter diffraction spikes, remains fixed throughout the study. The image represents data captured over a 50-second period prior to the occultation. The 10-pixel radius circle depicts the aperture used for measurements, and containing 80\% of the total flux.}
   \label{fig1}
   \end{figure*}

On December 12, 2023, we observed a single-chord occultation of Betelgeuse from the AstroCamp observatory\footnote{https://www.astrocamp.es/} in Nerpio, Southeast Spain, located just a few kilometers from the center of the occultation shadow path. We operated our SPAD camera in single-photon counting event mode for a total of 373.22 seconds starting at 01:12 UT. The expected time of Betelgeuse's occultation was at 01:15 UT. Figure~\ref{fig1} displays a defocus image of Betelgeuse, generated from an equivalent exposure time of 50 seconds
before the occulation by Leona. The right panel shows the $64 \times 64$ pixel image according to the arrangement of the SPAD sensor within each individual and adjacent pixels \citep[see][for more details]
{Ruben}. The inset shows a zoom into the arrangement of $2 \times 2$ pixels. The left panel shows a symbolic image created from
the $64 \times 64$ SPAD sensors. It depicts the Betelgeuse defocus signal as well as the fainter diffraction pattern caused by the telescope spider structure holding the secondary mirror. These X-shaped features are clearly visible in Figure~\ref{fig1}, left panel. We have applied a hot pixel\footnote{individual SPAD pixels that exhibit a higher-than-usual DCR or that generate spurious signals even in the absence of incident light.} mask to the image, obtained from dark count data collected over six minutes just before observing Betelgeuse. 

Our analysis focuses on counting photons within a circular aperture defined to 
encompass $80\%$ of the signal emitted from the defocused image of Betelgeuse. The aperture size avoids the fainter difraction spikes, and remains fixed throughout the analysis presented in this work, encompassing a total of 316 pixels on the detector (see Figure~\ref{fig1}, right panel). After masking the hot pixels, there are 304 pixels within the aperture and a total of 3777 pixels in the full array. We measured a median DCR of $13$ events per second ($1.3\times10^{-5}$ events per microsecond) per pixel. This rate is negligible compared to the signal from Betelgeuse.

In addition, the occultation of Betelgeuse was observed using a PlaneWave CDK 12.5-inch telescope, paired with a QHY-268-M CMOS camera and equipped with a Baader Bessel r photometric filter. To improve the accuracy of the data collected, a custom mask consisting of 8 sub-apertures, each of 4 cm in diameter, was used \cite[see][]{Sigismondi2023}. This dataset enabled a comparison with our SPAD observations. During the occultation campaign, the seeing conditions were about $2\arcsec$.

\section{Results}

\subsection{Description of the detected event statistics}

Addressing practical challenges in SPAD arrays such as optical cross-talk\footnote{spurious events from spill-over signal affecting neighboring pixels}, a limited number of pixels, and the impact of afterpulses\footnote{delayed secondary pulses following the detection of a primary photon}, we explore deviations from the ideal Poisson behavior of photon statistics. We use a binomial distribution to describe the probability of detecting $N$ events in a given time interval; this has been proven to describe well event statistics in SPAD arrays over both small and large photon limits \citep[e. g.][]{Houwink,Scholes}. In this work, we adopt a generalised negative binomial distribution (NBD) suited for our case study, which aligns with the theory of avalanches as demonstrated in \citet[][]{BR}, i.e.,

\begin{equation}
P(N) = \frac{(\lambda T / (1 + z\lambda T))^N}{N!} \left[\prod_{i=1}^{N-1} (1 + i \, z)\right] (1 + z \lambda T)^{-1/z} \, ,
\label{eqNBD}
\end{equation}

\noindent
where $\lambda$ is the mean number of events per unit time ($\mu \rm s$ in our case) and $T$ is the considered time interval. $z$ is related to the physical parameters of the SPAD array as follows:

\begin{equation}
z \lambda T =  x - \frac{\lambda T}{n_{\rm pix}^{\rm eff}}  \, ,
\label{eqz}
\end{equation}

\noindent
with $x$ representing the optical cross-talk and $n_{\rm pix}^{\rm eff}$ the number of effective pixels in the region of interest, i.e.,

\begin{equation}
\frac{{1}}{{n_{\rm pix}^{\rm eff}}} = \frac{{1}}{{2}}  \frac{{\sum_{i=1}^{n_{\rm pix}} N_i^{2}}}{{N_{\rm tot}^2}}  \, ,
\label{eqneff}
\end{equation}

\noindent
where $n_{\rm pix}$ is the number of pixels in the detector within the region of interest, $N_i$ represents the number of photons captured in the $i$-th pixel  during the specified time interval, and $N_{\rm tot} = \sum_{i=1}^{n_{\rm pix}} \, N_{\rm i}$. In the case of an ideal detector, where $z=0$, the NBD distribution given in Eq.~\ref{eqNBD} simplifies to the Poisson distribution.

\begin{figure*}[ht!]
  \centering
    \includegraphics[scale=0.51]{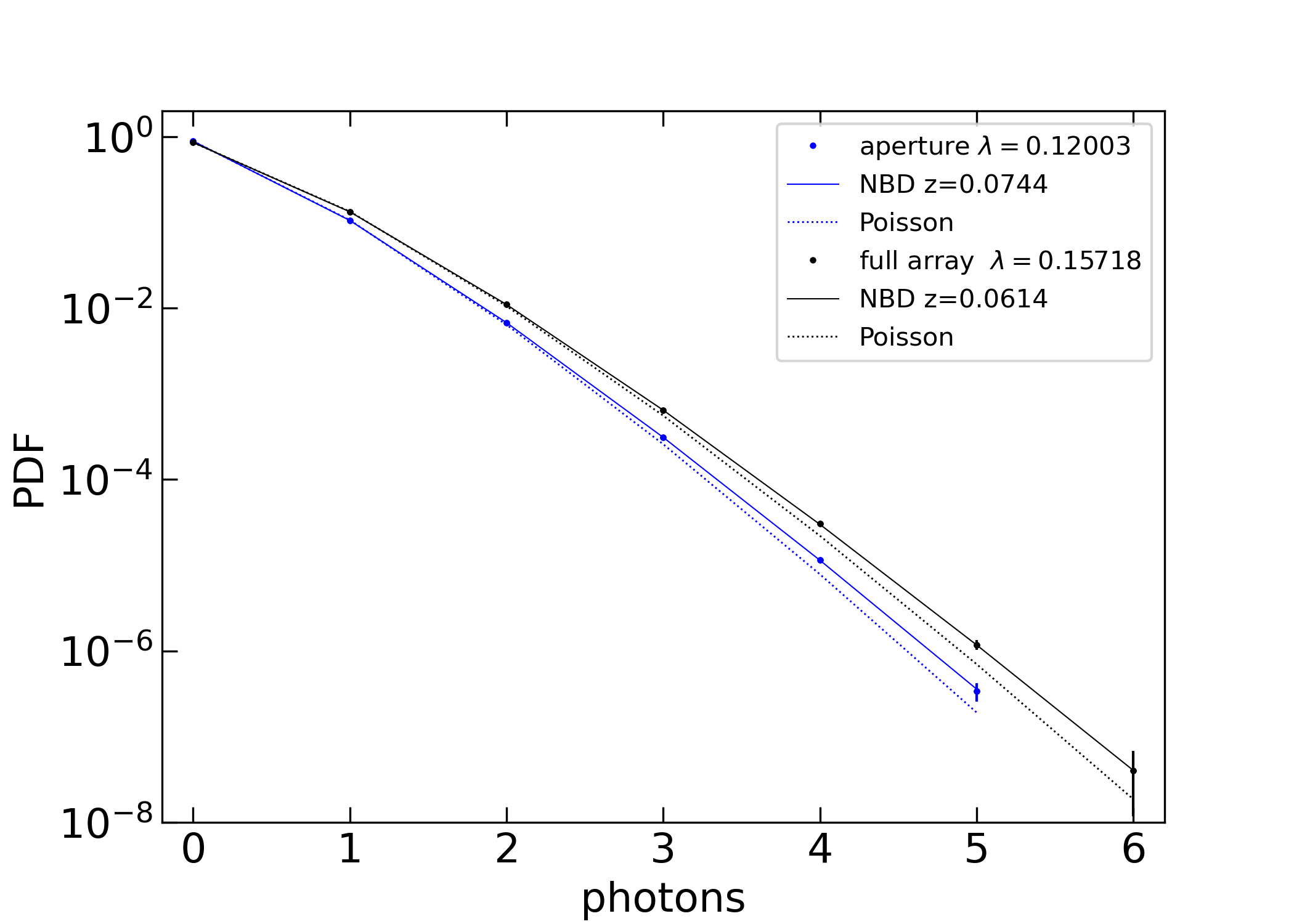}
    \includegraphics[scale=0.5]{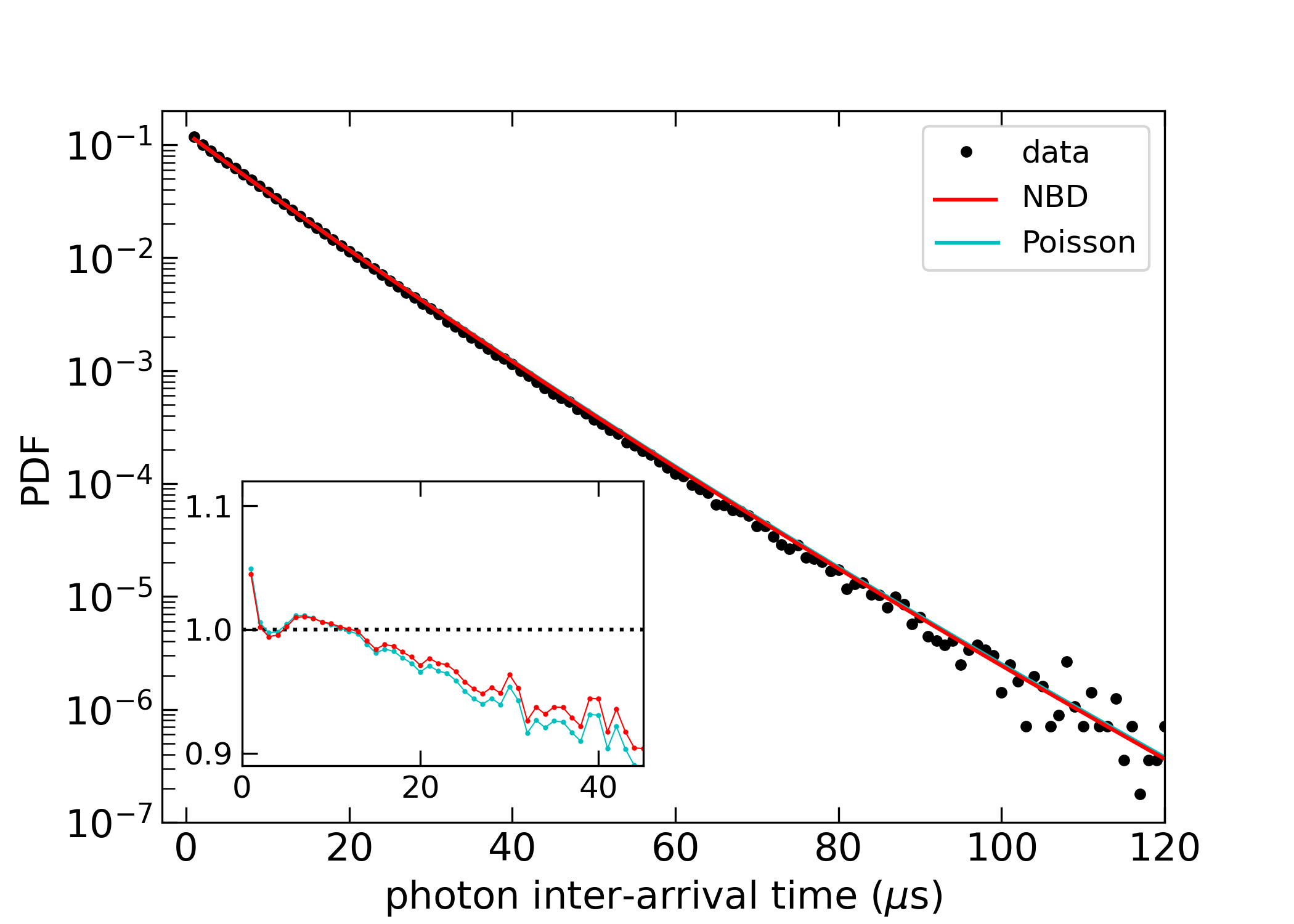}
    \caption{(left) Probability Distribution Function (PDF) prior to occultation, within the aperture (blue) and in the full array (black);
    dots with error bars represent measurements, while solid lines correspond to NBD fits to Eq. 1 for the indicated value of $z$. The Poisson results for the indicated values of $\lambda$ are shown as dotted lines. 
    (right) PDF of photon inter-arrival times; black dots represent the data within the aperture, while solid lines are the fits to Eq. 6 for the NBD (red) and for the Poisson distribution (cyan). The inset for $\Delta t<50\, \mu \rm s$ shows the effect of residual afterpulsing at $i=1$, and one can see beyond 10 s, one can observe the impact of fluctuations in $\lambda$ over scales of several  
    $\mu \rm s$, due to atmospheric short-time variability not included in our model.}
    \label{fig2}
\end{figure*}

The left panel in Figure~\ref{fig2} shows the probability distribution of $N$ detected events over a $1 \, \mu \rm s$ interval; counted within the circular aperture (blue dots and error bars) and collected in the full array (black dots and error bars). In total, there are $6,001,612$ events  within the circular aperture and $7,859,056$ photons in the full array. These data were recorded during the 50-second interval preceding the occultation by Leona. Error bars were estimated using the expression $\sqrt{(P(N)(1-P(N))}/\sqrt{50e6}$.

\begin{table*}[htbp]
\centering
\caption[]{Probability of detecting $N$  photons in a $1 \, \mu \rm s$ interval for Betelgeuse's circular aperture and for the full array, based on measurements from the 50 seconds preceding the occultation. Additionally, the table includes expected probabilities from best-fit NBD and Poisson models ($\chi^2$ are given for each case) as described in the text and showed in Figure~\ref{fig2}).}
\label{tabPN50s}
\[
\begin{array}{lccc|cccc}
\hline
 & \multicolumn{3}{c}{\text{Betelgeuse's aperture}} |& \multicolumn{3}{c}{\text{Full array}} \\
\hline
\noalign{\smallskip}
N & \multicolumn{1}{c}{\text{Data}} & \multicolumn{1}{c}{\text{NBD $$}} & \multicolumn{1}{c}{\text{Poisson}} |& \multicolumn{1}{c}{\text{Data}} & \multicolumn{1}{c}{\text{NBD }} & \multicolumn{1}{c}{\text{Poisson }}  \\
 & \multicolumn{1}{c}{\text{}} & \multicolumn{1}{c}{\text{$\chi^2=1.4$ $$}} & \multicolumn{1}{c}{\text{$\chi^2=1988$}} |& \multicolumn{1}{c}{\text{}} & \multicolumn{1}{c}{\text{$\chi^2=2.3$ }} & \multicolumn{1}{c}{\text{$\chi^2=2400$}}  \\
\hline
\noalign{\smallskip}
0  & 8.8739 \cdot 10^{-1} \pm 4.5 \cdot 10^{-5} & 8.8739 \cdot 10^{-1} & 8.8692 \cdot 10^{-1} & 8.5523 \cdot 10^{-1} \pm 5.0 \cdot 10^{-5} & 8.5526 \cdot 10^{-1}  & 8.5458 \cdot 10^{-1}  \\
1  & 1.0551 \cdot 10^{-1} \pm 4.3 \cdot 10^{-5} & 1.0551 \cdot 10^{-1} & 1.0640 \cdot 10^{-1} & 1.3307 \cdot 10^{-1} \pm 4.8 \cdot 10^{-5} & 1.3308 \cdot 10^{-1}  & 1.3426 \cdot 10^{-1} \\
2  & 6.7747 \cdot 10^{-3} \pm 1.2 \cdot 10^{-5} & 6.7684 \cdot 10^{-3} & 6.4102 \cdot 10^{-3} & 1.1027 \cdot 10^{-2} \pm 1.5 \cdot 10^{-5} & 1.1021 \cdot 10^{-2}  & 1.0578 \cdot 10^{-2} \\
3  & 3.0834 \cdot 10^{-4} \pm 2.5 \cdot 10^{-6} & 3.1081 \cdot 10^{-4} & 2.5857 \cdot 10^{-4} & 6.4192 \cdot 10^{-4} \pm 3.6 \cdot 10^{-6} & 6.4551 \cdot 10^{-4}  & 5.5720 \cdot 10^{-4} \\
4  & 1.1580 \cdot 10^{-5} \pm 4.8 \cdot 10^{-7} & 1.1447 \cdot 10^{-5} & 0.7857 \cdot 10^{-5} & 3.0840 \cdot 10^{-5} \pm 7.9 \cdot 10^{-7} & 2.9995 \cdot 10^{-5}  & 2.2079 \cdot 10^{-5} \\
5  & 3.4000 \cdot 10^{-7} \pm 8.2 \cdot 10^{-8} & 3.5929 \cdot 10^{-7} & 1.9182 \cdot 10^{-7} & 1.1800 \cdot 10^{-6} \pm 1.5 \cdot 10^{-7} & 1.1762 \cdot 10^{-6}  & 0.7020 \cdot 10^{-6} \\
6  & -                                          & -                    & -                    & 4.0000 \cdot 10^{-8} \pm 2.8 \cdot 10^{-8} & 0.4045 \cdot 10^{-7}  & 1.8657 \cdot 10^{-8} \\
\hline
\end{array}
\]
\end{table*}

The solid lines in Figure~\ref{fig2}, left panel, represent the fits to the probability distribution of the data for both the aperture (blue) and the full array (black). These fits were obtained by averaging over 50-second interval across an ensemble of NBD spanning 50 time intervals of 1 second each. This method accounts for fluctuation in the mean rate of photons, $\lambda_j$, due to the presence of clouds during observations. Specifically, $P(N) = (\sum_{j=1}^{50} \, P_j(N))/50 $, where $P_j(N)$, corresponding to each $\lambda_j$, is defined in Eq.~\ref{eqNBD}. As a result, we obtained the following values for the $z$ parameter: $z_{\rm ap} = 0.0744 \pm 0.0016$ for the aperture and $z_{\rm fa} = 0.0614 \pm 0.0013$ for the full array. Table~\ref{tabPN50s} lists the values of the photon probability distribution, including the data and NBD fit, for both the aperture and the full array.

Hence, we estimate the optical cross-talk using Eq.~\ref{eqz}, based on the mean number of photons in $1 \, \mu s$ and the effective number of pixels for both the aperture and the full array. For the aperture we have $\lambda_{ap} = 0.12003$ and $n_{\rm pix, ap}^{\rm eff} = 78.46$, and for the full array, $\lambda_{\rm fa} = 0.15718$ and $n_{\rm pix, fa}^{\rm eff} = 134.47$. As expected, we derive similar values for the optical cross-talk: $x_{\rm ap} = 1.046 \times 10^{-2} \pm 1.92 \, \times \, 10^{-4}$ for the aperture and $x_{\rm fa} = 1.0858 \times 10^{-2} \pm 2.04 \, \times \, 10^{-4}$ for the full array. These values, when combined, yield an optical cross-talk for our IMSE-2CASS SPAD array of $1.067 \times 10^{-2} \pm 1.70 \, \times \, 10^{-4}$. This result is consistent with measurements obtained from individual pixels conducted in the laboratory.

To assess the impact of residual afterpulse effects and test the consistency of our photon statistics model, we examine the inter-arrival time probability distribution, $P_{\rm it}$, of photons detected within Betelgeuse's circular aperture. $P_{\rm it}$ is expected to follow an exponential form  $\lambda e^{-\lambda \Delta{t}} \), with $\Delta{t}$ representing the time between consecutive events in a Poisson process with rate $\lambda$. This expectation arises from the inherent nature of the Poisson process that governs photon arrivals in an ideal SPAD detector. However, for a real SPAD array, $P_{\rm it}$ is more accurately described by a negative binomial distribution, as mentioned previously. The inter-arrival time probability distribution in this scenario is characterized by the expressions provided below,

\begin{equation}
P_{it}(\Delta{t}=0) \equiv P_{it}(i=0) =  1 - \frac{{1-P(0)}}{{\lambda}}  \,,
\label{eqPit0}
\end{equation}

\begin{equation}
P_{it}(\Delta{t}) \equiv P_{it}(i) =  (1 - P(0)) \, P(0)^{i-1} = \frac{{1 - (1 + \lambda \, z)^{-1/z}}}{{(1 + \lambda \, z)^{(i-1)/z}}} \,.
\label{eqPit}
\end{equation}

\noindent
where $i$ is the inter-arrival time $\Delta{t}$ in $\mu\text{s}$.

Figure~\ref{fig2}, right panel, displays the inter-arrival time distribution derived from a sample of 5,630,240 photons with $\Delta{t}>0$ within Betelgeuse's circular aperture, collected in the $50-second$ interval preceding the occultation. The red line represents the NBD distribution according to Eq.\ref{eqPit}, using the $z_{\rm ap}$ parameter obtained from fitting the photon probability distribution (see Figure~\ref{fig2}, left panel). The Poisson distribution is represented by the cyan line. 

Note that the models results in Figure~\ref{fig2} are obtained from Eqs.~\ref{eqPit0} and~\ref{eqPit}, computed over a large number of time segments, $n$, of the total observation time (50 s), each using the locally estimated value of $\lambda$. The choice of $n$ presents a dilemma: larger values of $n$ better account for fluctuations due to atmospheric short-time variability, but they also increase the impact of spurious, essentially Poissonian, fluctuations. Ideally, if data and model agreement could be achieved with an $\lambda$ small enough that Poissonian fluctuations remain negligible, this would not pose a concern. However, this was not achievable. Therefore, we had to correct for these effects in the local estimates of $\lambda$. With these corrections, the model's predictions should converge to accuracy as $n$ increases. We have evaluated the predictions for $\lambda$ values of 10,000 and 50,000 and observed negligible differences; the results for the latter are depicted in Figure~\ref{fig2}.

The discrepancy in the value of $P(i=1)$ can be attributed to residual afterpulse effects on the $\mu s$ time scale). Under the following assumptions, we can easyly provide a simple model to account for this effect: a SPAD pixel activated in a given microsecond has a probability $P_0$ of inducing a spurious detection in the same pixel in the next microsecond, due to an avalanche lasting more than a microsecond. The probability of a spurious detection in subsequent microseconds is zero. With this hypothesis, for $i=1$, we have:
\[ P(i=1) = P_0 + (1-P_0)(1 - P(0)), \]

\noindent 
and for $i > 1$,

\[ P(i) = (1-P_0)(1-P(0))P(0)^{i-1}. \]

\noindent 
To fit the data, we require $P_0 = 5.9 \times 10^{-3}$. Despite adjustments, the data still falls below model predictions beyond 10s,
likely driven by fluctuations of several $\mu\text{s}$ in $\lambda$ caused by short-time atmospheric variability, which, as mentioned earlier, is unaccounted for in the model.

\subsection{Lightcurve of Betelgeuse's occultation by Leona}

\begin{figure*}
  \centering
  \includegraphics[scale=0.4]{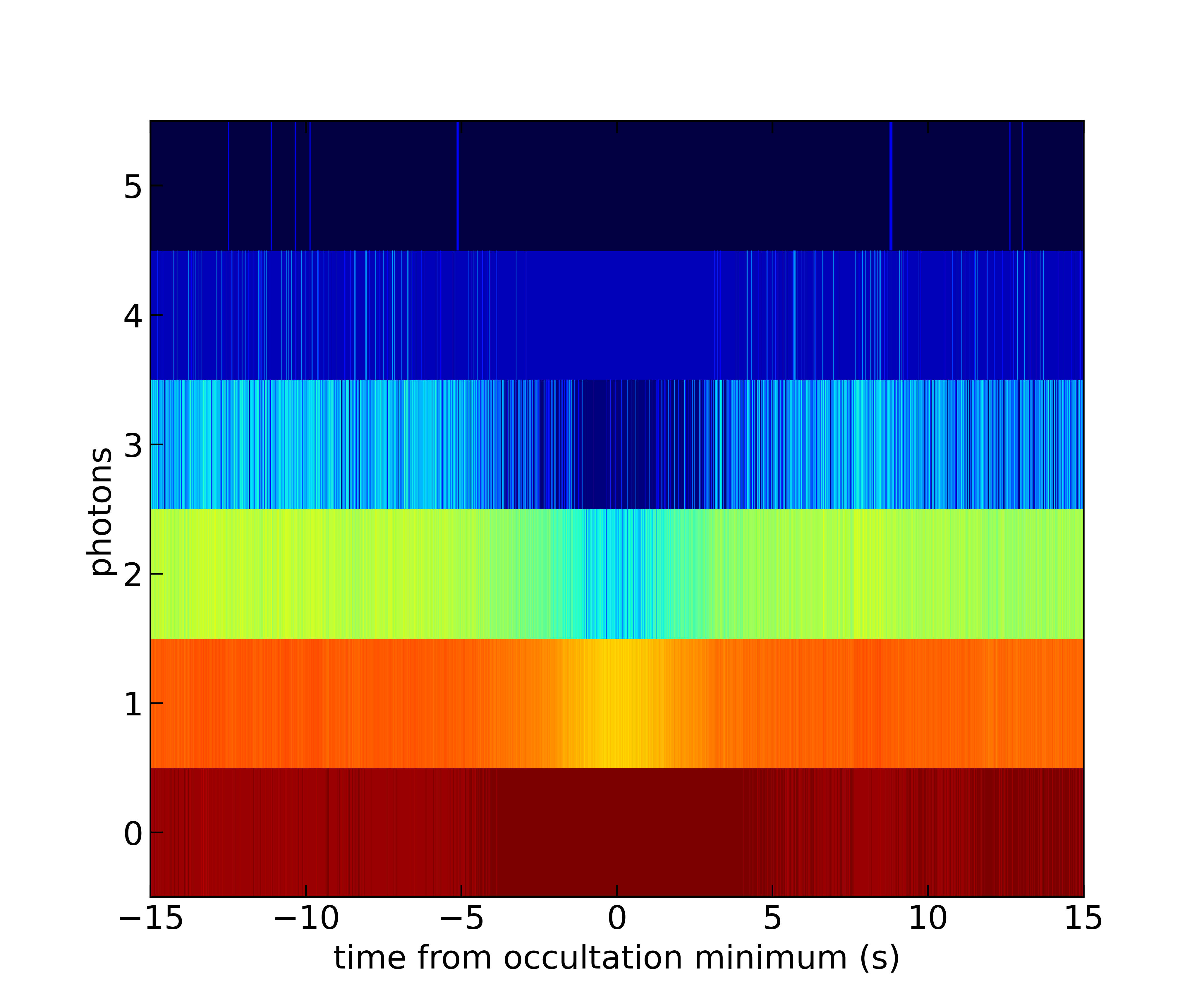}
  \includegraphics[scale=0.55]{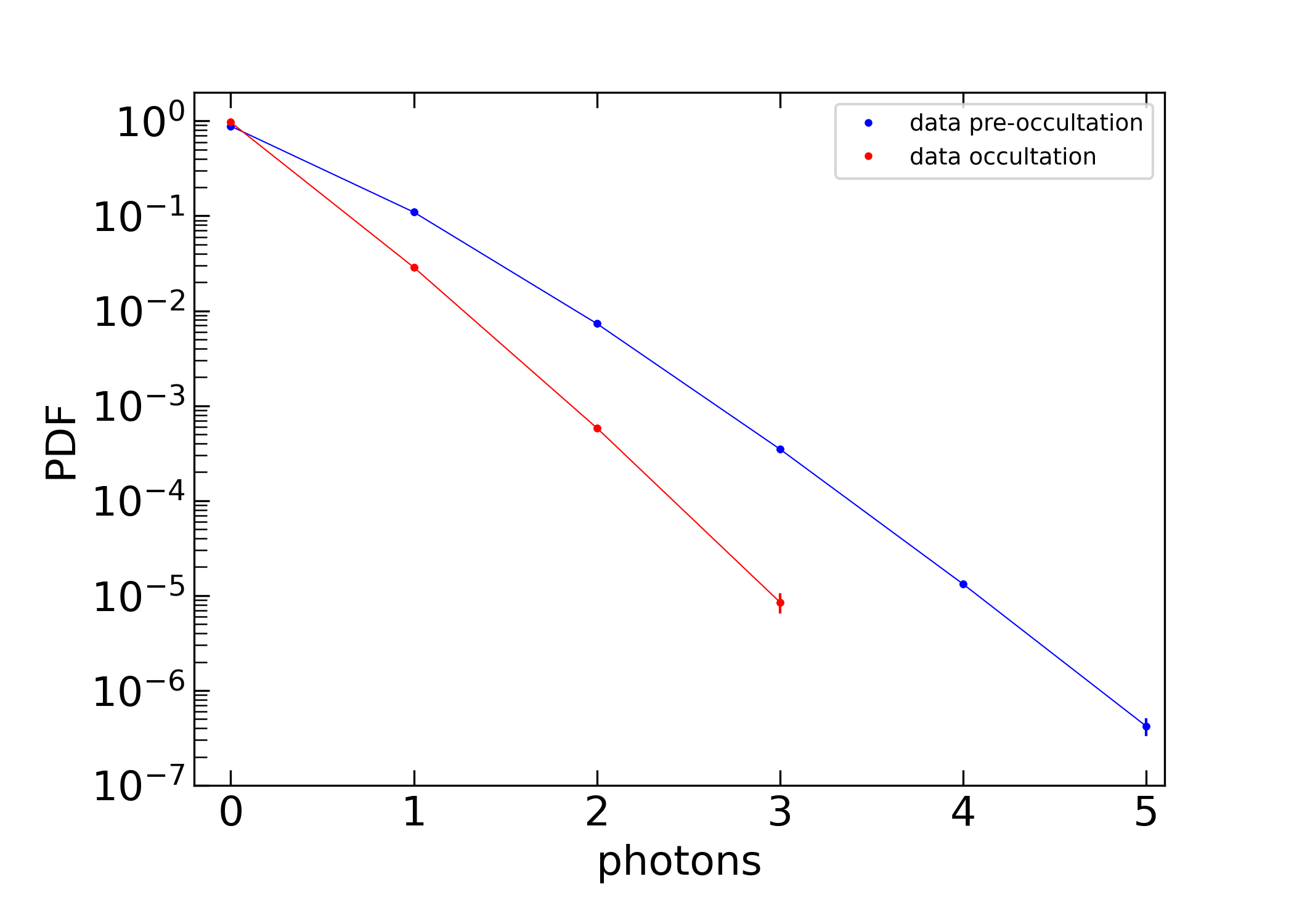}
  \caption{(Left): Photon data from Betelgeuse $\pm 15 s$ around the peak of the occultation, obtained with our SPAD-based camera with $1 \, \mu \rm s$ time resolution. (Right:) Dots with error bars represent the PDF within the aperture prior to (blue) and during (red) occultation; note that the full lines are included for clarity but do not represent any fit or prediction.}
  \label{fig:LCs}
\end{figure*}

Stellar occultations occur when an asteroid passes in front of a distant star, briefly obscuring its light and causing an intensity drop. This phenomenon is depicted in Figure~\ref{fig:LCs} for the occultation of Betelgeuse by Leona. The left panel displays the number of photons recorded every microsecond within Betelgeuse's circular aperture during a $\pm15$ s interval centered on the time of the lowest detected photon count, coinciding with the peak of the occultation. Figure~\ref{fig:LCs}, right panel, compares the photon probability distribution over $1 \, \mu \rm s$ for the 50-s data recorded prior to the start of the occultation (blue points with error bars) with that from the $\pm1$ s data around the maximum of the occultation (red points with error bars). The ratio of the minimum intensity observed during the occultation to the intensity outside the occultation window is determined from the ratio of the mean number of photons detected at the peak to those detected prior to the occultation, i.e. $0.02696/0.1213 = 0.2222$. This corresponds to a drop of $77.78\%$ in Betelgeuse's intensity.

\begin{figure}
  \includegraphics[scale=0.55]{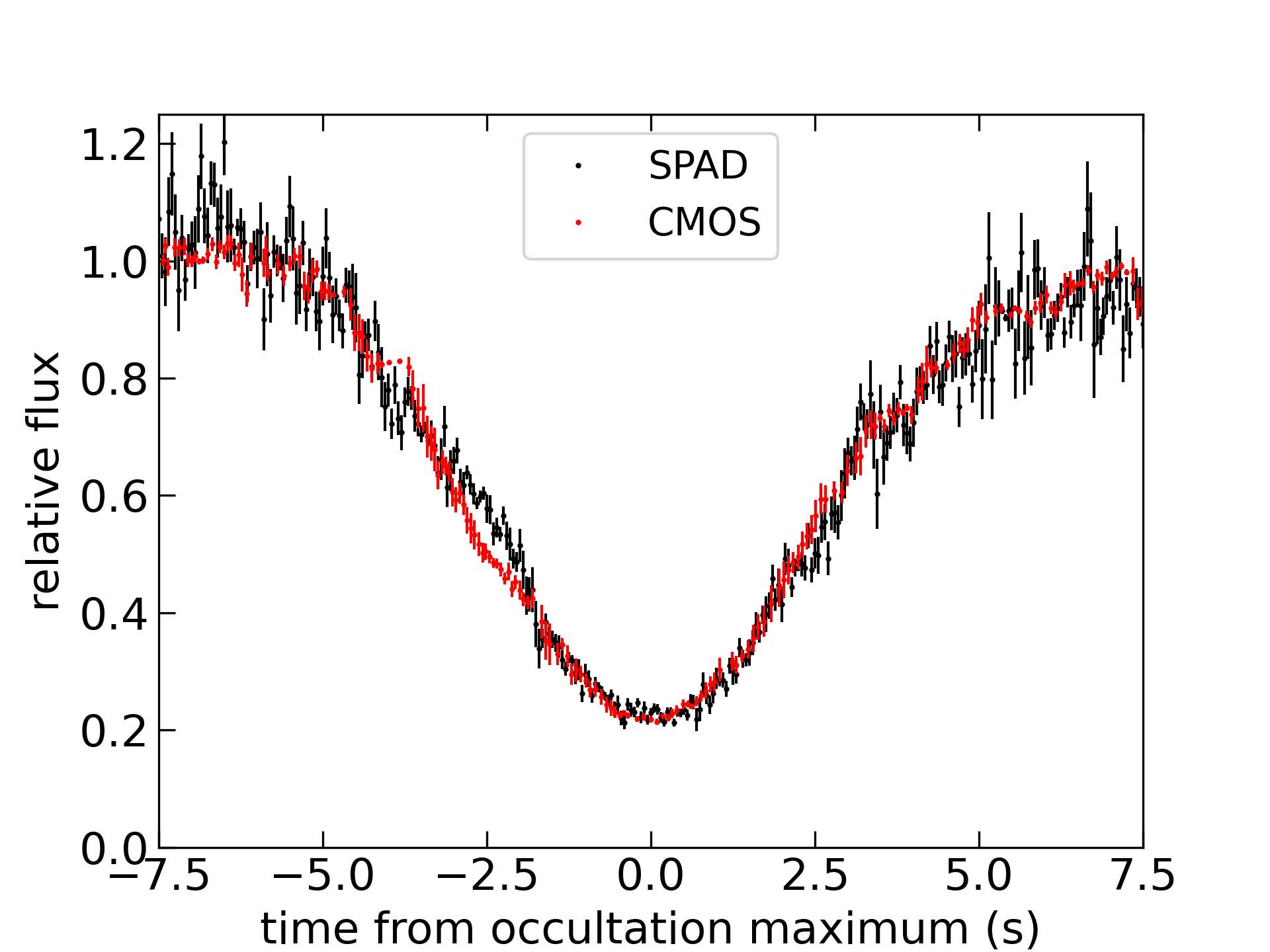}
  \caption{Good agreement of light curves derived from our observations of Betelgeuse's occultation using our SPAD detector (black) and from an independent CMOS detector with exposure frames of 50 ms (red). The CMOS exposures exhibit a signal-to-noise ratio that is at least 50 times higher than that of the SPAD, primarily due to the low $3.5\%$ fill factor of the SPAD detector.}
  \label{fig:Cantalapiedra}
\end{figure}

The method presented herein relies on several simplified assumptions to derive the angular size of Betelgeuse during the occultation event by Leona. Specifically, we assume that both the star and the asteroid are perfect uniform disks, a model that facilitates straightforward geometric analysis. Limb darkening is initially neglected to simplify the relationship between the observed intensity drop and the fraction of the star's disk obscured by the asteroid. Under these assumptions, we explore the relationship between the  intensity drop observed during an occultation event and the relative angular sizes of the celestial bodies involved. The parameters of the model include:

   \[
      \begin{array}{lp{0.8\linewidth}}
         \theta_{\text{s}}  & angular size of the star,                      \\
         d_{\text{ast}}  & distance from the observer to the asteroid,                       \\
         D_{\text{ast}}  & physical size of the asteroid,                       \\
         \theta_{\text{ast}}  & angular size of the asteroid,                      \\
         I_{\text{drop}} & intensity drop during the occultation.
      \end{array}
   \]
\noindent

The intensity drop can be expressed as follows:

\begin{equation}
I_{\text{drop}} = 1 - \frac{A_{\text{covered}}}{A_\text{s}} = 1 - \frac{A_{\text{ast}}}{A_\text{s}} = 1 -  \left(\frac{\theta_\text{ast}}{\theta_\text{s}}\right)^2 \, ;
\label{eq:Idrop}
\end{equation}
\noindent
where \(A_{\text{covered}}\) represents the area of the star covered by the asteroid, and \(A_s\) denotes the total area of the star. Assuming both bodies are viewed as uniform disks, \(A_{\text{covered}}\) closely approximates the area of the asteroid. The proportionality between their areas corresponds to their angular sizes, due to the significant distances of both bodies from the observer. Hence, the angular size of the star (\(\theta_{s}\)) relative to the angular size of the asteroid (\(\theta_\text{ast}\)) can be expressed using the  intensity drop ($I_{\text{drop}}$), as follows,

\begin{equation}
\theta_s = \theta_\text{ast} \left( \frac{1}{\sqrt{1 - I_{\text{drop}}}} \right) \, .
\label{eq:angdstar}
\end{equation}

Using the distance and physical size of the asteroid, we calculate its angular size with the formula: $\theta_{\text{ast}} = 2 \arctan\left(D_{\text{ast}}/2 \, d_{\text{ast}}\right)$. For Leona, with a distance of $1.8008$ AU and an effective diameter of $66 \pm 2$ km as reported by \citet{Ortiz}, its angular size is 50.5 mas. This calculation allows the direct estimation of Betelgeuse's angular diameter by incorporating $\theta_{\text{ast}}$ and the observed intensity drop $I_{\text{drop}}$ during its occultation. As a result, we estimate Betelgeuse's angular diameter to be 57.26 mas in the SDSS $g'$ visible band, in accordance with Eq.~\ref{eq:angdstar}. 

Incorporating limb darkening into the occultation model improves the accuracy of measuring the star's angular diameter. This approach accounts for the reduced brightness at the edges of the stellar disk, indicating that a more significant portion of the star contributes to the observed intensity drop during occultation, leading to a larger estimated diameter. This is particularly relevant for red supergiant stars like Betelgeuse, where limb darkening is quite significant \citep[see e.g.][]{Neilson,Claret}. However, addressing these corrections falls beyond the scope of this paper.

In our observations of Betelgeuse's occultation, we used two detectors systems: a SPAD detector and a CMOS detector. Their respective light curves are shown in Figure~\ref{fig:Cantalapiedra}, with the SPAD light curve in black and the CMOS in red, both using 50-ms exposures. Despite their different sensitivities, both detectors showed a noteworthy agreement in the light curves. The CMOS detector exhibited a signal-to-noise ratio about 50 times higher than that of the SPAD, attributed to its $100\%$ fill factor which significantly enhances light collection efficiency. In contrast, the SPAD detector is limited by a low $3.5\%$ fill factor.



\section{Conclusions}

We report the observation of Betelgeuse's occultation by asteroid (319) Leona on December 12, 2023, using a 64×64 pixel SPAD array on a 10-inch telescope at AstroCamp Observatory in Spain. This marks a significant advancement in using SPAD technology in astronomy, achieving 1-microsecond temporal resolution in the light curve observations of Betelgeuse in visible wavelengths. We addressed SPAD-specific challenges by adopting a generalized negative binomial distribution for a more accurate data description, finding $1.07\%$ optical cross-talk and minimal impact from afterpulses. It is crucial to recognize that using the NBD to describe our data statistics is not just a convenient model for deviations from Poissonian distribution but a necessity arising from the underlying physics of cross-talk and the limited number of pixels. it has been rigorously demonstrated that the resulting probability distribution must be the NBD, with the additional parameter precisely defined by these hypotheses. This rigorous foundation allows us to accurately estimate the magnitude of cross-talk from small deviations from the Poisson distribution. Overall, this study highlights the potential of SPAD detectors for precise astronomical observations and provides a valuable dataset for the community.

The observation of Betelgeuse's intensity dropped enable us to measure its angular diameter at 57.26 mas in the SDSS g-band using a simplified uniform disk model and the known properties of Leona. Accurately determining the angular diameter of red supergiants like Betelgeuse involves corrections for limb darkening, while the presence of hot spots and circumstellar materials further complicates diameter estimations. Addressing these corrections falls beyond the scope of this paper. Moreover, finding Betelgeuse's diameter can be achieved by fitting the entire lightcurve data with a detailed occultation model, following the approach in previous studies of other stellar occultations using specific tools such as SORA\footnote{https://sora.readthedocs.io/en/latest/index.html} and  Occult\footnote{http://lunar-occultations.com/iota/occult4.htm}. We are making the raw data taken with our SPAD camera, and the light curve with $1$ ms time resolution publicly available to contribute to that task with the highest time resolution observations obtained during Betelgeuse's occultation by Leona.\footnote{Electronic tables will be made available at CDS.}

\citet{Gilliland} resolved Betelgeuse's photosphere spatially using the HST Faint Object Camera, finding a UV diameter of $108 \pm 4$ mas, which is 1.9 times the uniform disk diameter found in this work in the visible at 480 nm (central wavelength of the SDSS-g' filter) from Betelgeuse's occultation by Leona and 2.6 larger than the $41.01 \pm 0.41$ uniform disk diameter obtained using near-IR VLT interferometry \citep{Montarges2014}. These larger UV and visible diameters suggest the presence of an extended chromosphere, akin to the Sun's hot temperature inversion layer \citep[see][]{Wheeler}.

\begin{acknowledgements}
We acknowledge Agust\'in S\'anchez and Fernando \'Abalos for providing access and support at the AstroCamp Observatory, and Ernesto S\'anchez Blanco for helping with the optics set-up of the SPAD array. FP thanks Luca Izzo, Luca Zampieri, Sergio J. Fern\'andez Acosta, H\'ector De Paz Mart\'in, Edward Schlafly, Rosario Cosentino and Luca Di Fabrizio for fruitful discussions. FP, RGM, EP, JALB, ARV and GdR thank the support from the Spanish MICINN funding grant PDC2023-145909-I00. FP and EP acknowledge financial support from the Severo Ochoa grant CEX2021-001131-S funded by MCIN/AEI/ 10.13039/501100011033. RGM thanks the support from the Spanish Government through Ayudas para la Formación del Profesorado Universitario (FPU) under Grant FPU19/03410.
\end{acknowledgements}


\begin{thebibliography}{}

\bibitem[Acconcia et al.(2023)]{Acconcia} Acconcia, G., Ceccarelli, F., Gulinatti, A., et al.\ 2023, Optics Express, 31, 33963
\bibitem[Barbieri et al.(2007)]{Barbieri} Barbieri, C., Dravins, D., Occhipinti, T., et al.\ 2007, Journal of Modern Optics, 54, 191
\bibitem[Betancort-Rijo(2000)]{BR} Betancort-Rijo, J.\ 2000, Journal of Statistical Physics, Vol. 98, Nos. 3/4
\bibitem[Boahen(2000)]{Boahen} Boahen, K. A. \ 2000  in IEEE Transactions on Circuits and Systems II: Analog and Digital Signal Processing, vol. 47, no. 5, 416
\bibitem[Bonanno et al.(2016)]{Bonanno} Bonanno, G., Belluso, M., Zappa, F. et al. \  2005, Exp Astron 19, 163–168
\bibitem[Bronzi et al.(2016)]{Bronzi} Bronzi, D., Villa, F., Tisa, S., et al.\ 2016, IEEE Sensors Journal, 16, 3
\bibitem[Cosentino et al.(2004)]{Cosentino} Cosentino, R., Bonnano, G., Belluso, M., et al. \ 2004, in Scientific Detectors for Astronomy, The Beginning of a New Era, Astrophysics and Space Science Library, vol 300, pag 29
\bibitem[Cannon et al.(2023)]{Cannon} Cannon, E., Montarg{\`e}s, M., de Koter, A., et al.\ 2023, \aap, 675, A46. 
\bibitem[Claret \& Bloemen(2011)]{Claret} Claret, A. \& Bloemen, S.\ 2011, \aap, 529, A75
\bibitem[Cusini et al.(2022)]{Cusini} Cusini, I., Berretta, D., Conca, E., et al.\ 2022, Frontiers in Physics, 10, 906671
\bibitem[Dolan et al.(2016)]{Dolan} Dolan, M.~M., Mathews, G.~J., Lam, D.~D., et al.\ 2016, \apj, 819, 7
\bibitem[Drevon et al.(2024)]{Drevon} Drevon, J., Millour, F., Cruzal{\`e}bes, P., et al.\ 2024, \mnras, 527, L88
\bibitem[Eisenhauer et al.(2023)]{Eisenhauer} Eisenhauer, F., Monnier, J.~D., \& Pfuhl, O.\ 2023, \araa, 61, 237
\bibitem[Erdogan et al.(2022)]{Erdogan} Erdogan, A. T., Al Abbas, T., Finlaysonet, N., al.\ 2022, IEEE Journal of Solid-State Circuits, vol. 57, no. 6, 1649
Haubois
\bibitem[Gilliland \& Dupree(1996)]{Gilliland} Gilliland, R.~L. \& Dupree, A.~K.\ 1996, \apjl, 463, L29
\bibitem[Gomes-J{\'u}nior et al.(2022)]{SORA} Gomes-J{\'u}nior, A.~R., Morgado, B.~E., Benedetti-Rossi, G., et al.\ 2022, \mnras, 511, 1167. 
\bibitem[Gomez-Merchan et al.(2023)]{Ruben} Gomez-Merchan, R., Leñero-Bardallo, J. A., Rosa-Vidal, R. d. L., \& Rodr\' iguez-V\'azquez, \'A.\ 2023, ESSCIRC 2023-IEEE 49th European Solid State Circuits Conference (ESSCIRC), Lisbon, Portugal
\bibitem[Hadfield et al.(2023)]{Hadfield} Hadfield, R.~H., Leach, J., Fleming, F., et al.\ 2023, Optica, 10, 1124
\bibitem[Houwink et al.(2021)]{Houwink} Houwink, Q., et al.\ 2021, Opt. Express 29, 39920
\bibitem[Huang et al.(2023)]{Huang} Huang, H. -H., Huang, T. -Y., Liu, C. -H., et al.\ 2023, IEEE Sensors Journal, vol. 23, no. 17, 19272
\bibitem[Matthews et al.(2022)]{Matthews} Matthews, N., Rivet, J.-P., Hugbart, M., et al.\ 2022, \procspie, 12183, 121830
\bibitem[Montarg{\`e}s et al.(2014)]{Montarges2014} Montarg{\`e}s, M., Kervella, P., Perrin, G., et al.\ 2014, \aap, 572, A17
\bibitem[Montarg{\`e}s et al.(2021)]{Montarges} Montarg{\`e}s, M., Cannon, E., Lagadec, E., et al.\ 2021, \nat, 594, 365
\bibitem[Morimoto et al.(2021)]{Morimoto} K. Morimoto et al.\ 2021, IEEE International Electron Devices Meeting (IEDM), San Francisco, CA, USA, 20.2.1
\bibitem[Neilson et al.(2011)]{Neilson} Neilson, H.~R., Lester, J.~B., \& Haubois, X.\ 2011, 9th Pacific Rim Conference on Stellar Astrophysics, 451, 117
\bibitem[Ortiz et al.(2024)]{Ortiz} Ortiz, J.~L., Kretlow, M., Schnabel, C., et al.\ 2024, \mnras, 528, L139
\bibitem[Scholes et al.(2023)]{Scholes} Scholes, S., Mora-Mart{\'\i}n, G., Zhu, F., et al.\ 2023, Scientific Reports, 13, 176
\bibitem[Sigismondiet al.(2023)]{Sigismondi2023} Sigismondi, C., Costa, C., \& Noschese, A. \ 2023, The Astronmer's telegram, 16374
\bibitem[Sigismondi et al.(2024)]{Sigismondi} Sigismondi, C., Costa, C., Noschese, A., et al.\ 2024, Journal for Occultation Astronomy, 14, 27
\bibitem[Townes et al.(2009)]{Townes} Townes, C.~H., Wishnow, E.~H., Hale, D.~D.~S., et al.\ 2009, \apjl, 697, L127
\bibitem[Vornicu et al.(2021)]{Vornicu} Vornicu, I., L\'opez-Mart\'inez, J. M., Bandi, F. N., Gal\'an, R. C., Rodr\'iguez-V\'azquez, \'A.\ 2021, IEEE Sensors Journal, vol. 21, no. 4, pp. 4776
\bibitem[Wayne et al.(2024)]{Wayne} Wayne, M., Ulku, A., Ardelean, A., et al.\ 2022, IEEE Transactions on Electron Devices, vol. 69, no. 6, 2865
\bibitem[Weiss et al.(2018)]{Weiss} Weiss, S.~A., Rupert, J.~D., \& Horch, E.~P.\ 2018, \procspie, 10701, 107010
\bibitem[Wheeler \& Chatzopoulos(2023)]{Wheeler} Wheeler, J.~C. \& Chatzopoulos, E.\ 2023, Astronomy and Geophysics, 64, 3.11
\bibitem[White et al.(2018)]{White} White, T.~R., Huber, D., Mann, A.~W., et al.\ 2018, \mnras, 477, 4403
\bibitem[Young et al.(2000)]{Young} Young, J.~S., Baldwin, J.~E., Boysen, R.~C., et al.\ 2000, \mnras, 315, 635
\bibitem[Zampieri et al.(2015)]{Zampieri2015} Zampieri, L., Naletto, G., Barbieri, C., et al.\ 2015, \procspie, 9504, 95040.
\bibitem[Zampieri et al.(2021)]{Zampieri2021} Zampieri, L., Naletto, G., Burtovoi, A., et al.\ 2021, \mnras, 506, 1585


\end{thebibliography}
\end{document}